\documentclass[epjCONF,columns]{svjour} 
\usepackage{graphicx}
\usepackage[varg]{txfonts} 
\usepackage[latin1]{inputenc}
\session-title{2011 Hadron Collider Physics Symposium}
\begin{document}
\title{Searches for High Mass Resonances and Exotics at the Tevatron}
\author{Ulrich Heintz\inst{1}\fnmsep\thanks{\email{ulrich\_heintz@brown.edu}} for the CDF and D0 Collaborations}
\institute{Brown University, 182 Hope Street, Providence, RI 02912, USA}
\abstract{I review recent searches for physics beyond the standard model from the CDF and D0 experiments at the Fermilab Tevatron collider, covering searches for fourth generation quarks, exotic resonances,  universal extra dimensions, and dark matter particles.} 
\maketitle
\section{The Tevatron at Fermilab}
\label{intro}

Run II of the Tevatron at Fermilab lasted from March 2001 to September 2011. During this run the Tevatron collider delivered $p\bar{p}$ collisions with a center of mass energy of 1.96~TeV and an integrated luminosity of about 12 fb$^{-1}$ to the two experiments, CDF and D0. 

\section{A Fourth Generation of Fermions}
\label{sec:1}

The standard model describes three fermion generations. It would be a straight forward extension of the standard model to add an additional generation by adding two fer\-mions each to the quark and lepton sectors. These quark and lepton fields would each consist of a left-handed weak doublet and two right-handed singlets. They would add new masses and mixing parameters to the model. Elec\-tro\-weak measurements constrain these parameters. The mass of the neutrino would have to be larger than half the $Z$ boson mass\cite{LEP} and the mass splitting between the two quarks, $t'$ and $b'$, would have to be less than the $W$ boson mass\cite{Gfitter}. The latter means that the decay of the heavier quark of the fourth generation to its lighter partner is kinematically suppressed and both fourth generation quarks must decay to third generation quarks via $t'\rightarrow bW$ and $b'\rightarrow tW\rightarrow bWW$, assuming moderate mixing exists between the third and fourth generations.

\subsection{The Search for a $t'$ Quark}

\begin{figure}[bh]
\centering
\includegraphics[width=0.9\columnwidth]{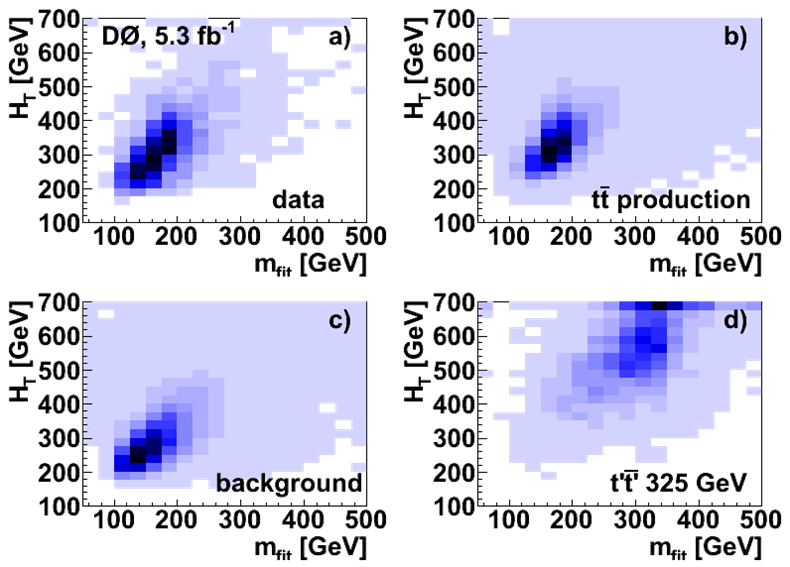}
\caption{Distributions of $H_T$ vs $m_{fit}$ for (a) data, (b) $t\bar{t}$ production, (c) other backgrounds, and (d) $t'\bar{t'}$ signal from the D0 experiment.}
\label{fig:HTvsMfit}       

\centering
\includegraphics[width=0.85\columnwidth]{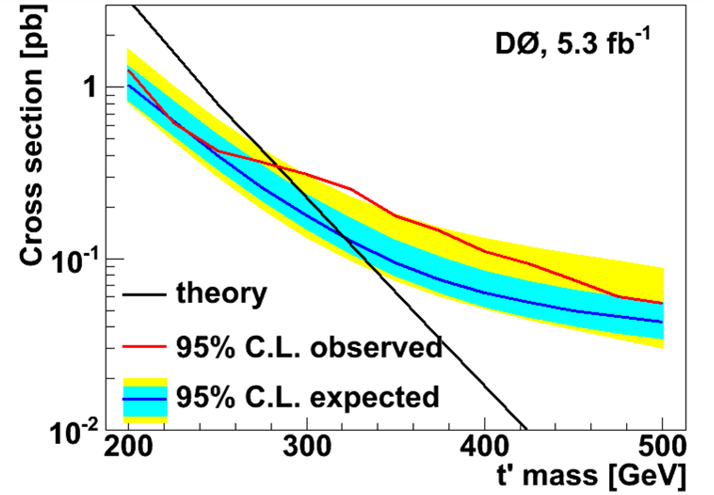}
\caption{Limits at 95\% C.L. for the $t'\bar{t'}$ production cross section from D0 without a $b$-tagging requirement.}
\label{fig:tplimitD0}       
\end{figure}

\begin{figure}[th]
\centering
\includegraphics[width=0.85\columnwidth]{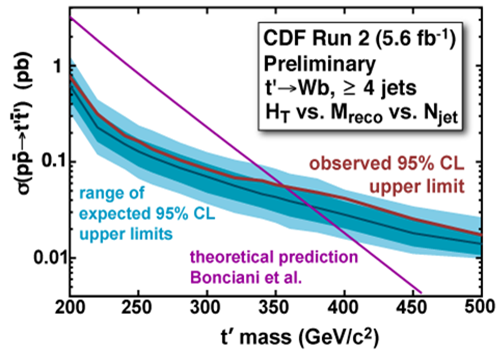}
\caption{Limits at 95\% C.L. for the $t'\bar{t'}$ production cross section from CDF requiring a $b$-tagged jet.}
\label{fig:tplimitCDF}       
\end{figure}

Both CDF and D0 have searched for $t'$ quarks that are pair-produced in $p\bar{p}$ collisions and decay to $bW$ \cite{tpsearches}. The most promising decay mode is the lepton+jets channel in which one $W$ boson decays to leptons ($e\nu$ or $\mu\nu$) and the other $W$ boson decays to jets. The final state thus is the same as for $t\bar{t}$ decay. Both experiments select events with a high-$p_T$ electron or muon, missing $p_T$, and at least four jets. The dominant background is $t\bar{t}$ production with a smaller contribution from $W$+jets production and other electroweak processes. As requiring $b$-tagged jets in the final state does not reduce the $t\bar{t}$ background significantly, D0 does not require any $b$-tags in the final state to make the search sensitive to any pair-produced objects that decay to $W$+jet. CDF performs two analyses, one without requiring a $b$-tagged jet and one requiring at least one $b$-tagged jet. The $t'\bar{t'}$ signal can be distinguished from $t\bar{t}$ production through the larger mass of the $t'$ quark. Thus both experiments perform a kinematic fit to the $t'\bar{t'}\rightarrow b\ell\nu \bar{b}q\bar{q}$ hypothesis and use the two-dimensional distribution of the best value for the mass of the $t'$ quark from the kinematic fit, $m_{fit}$, and the scalar sum of the momenta of all reconstructed objects, $H_T$, to search for a signal (Fig.~\ref{fig:HTvsMfit}). Figures~\ref{fig:tplimitD0} and \ref{fig:tplimitCDF} show the limits obtained for the $t'\bar{t'}$ production cross section. Thus $t'$ quarks that decay exclusively to a $W$ boson and a $b$-jet are disfavored for masses below 358~GeV.

\subsection{The Search for a $b'$ Quark}

\begin{figure}[hb]
\includegraphics[width=\columnwidth]{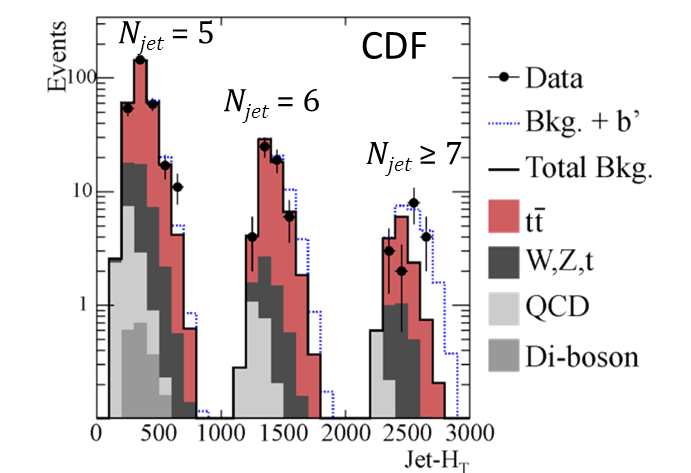}
\caption{Distribution of $H_T+1000(min(N_{jet},7)-5)$ from CDF.}
\label{fig:JetHT}       

\includegraphics[width=\columnwidth]{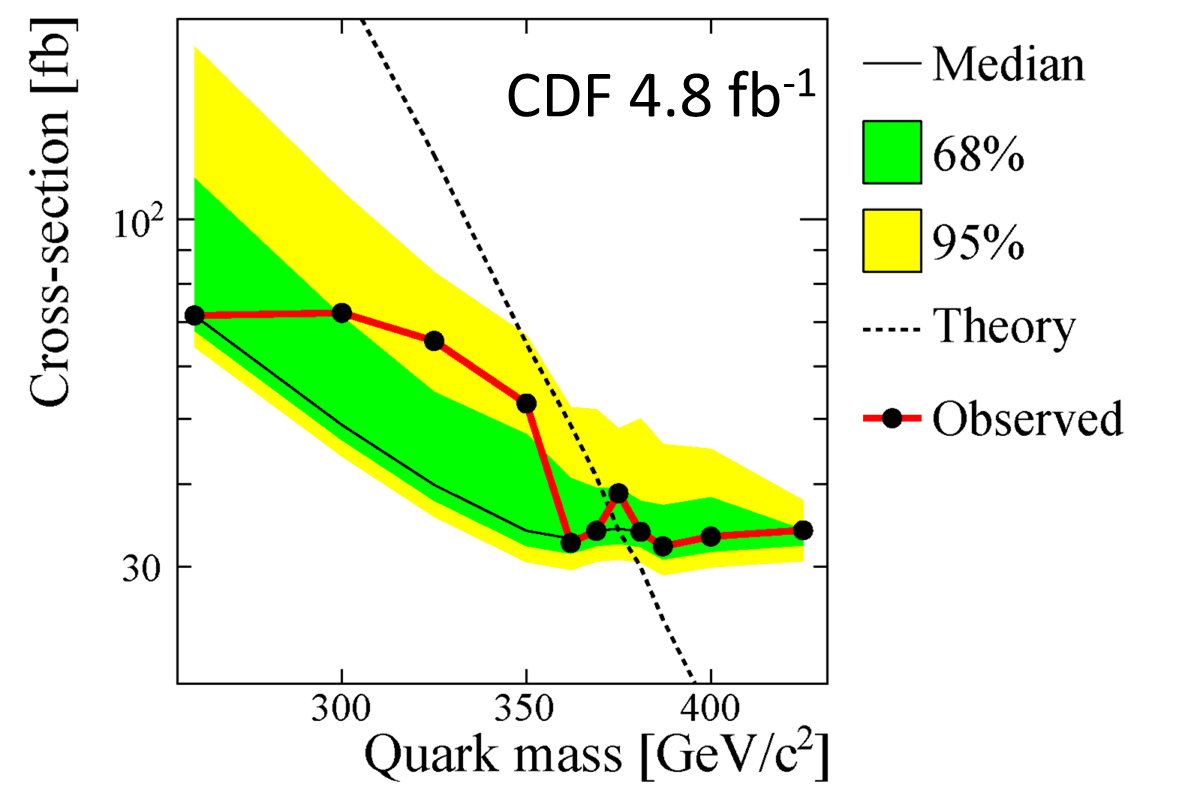}
\caption{Limits at 95\% C.L. for the $b'\bar{b'}$ production cross section from CDF.}
\label{fig:bplimits}       
\end{figure}

CDF has searched for $b'\bar{b'}$ production followed by the decay $b'\rightarrow tW$ \cite{bpsearch}. The final state contains four $W$ bosons and two $b$-jets. CDF selects events requiring a high-$p_T$ electron or muon, missing $p_T$, and at least five jets, one of which must be tagged as a $b$-jet. The $H_T$ distributions for events with different jet multiplicities $N_{jet}$ are then used to search for the signal, which is expected to have higher jet multiplicity and higher $H_T$ than the background, which is mostly from $t\bar{t}$ and $W$+jets production (Fig.~\ref{fig:JetHT}). Figure~\ref{fig:bplimits} shows the limits obtained for the $b'\bar{b'}$ production cross section. They disfavor $b'$ quarks for masses below 372~GeV.

\subsection{Heavy Neutrinos}

The neutrino of a fourth generation must be heavy to evade the limit on light neutrino flavors from the width of the $Z$ boson\cite{LEP}. If this heavy neutrino is a superposition of a Dirac and a Majorana particle there would be two mass eigenstates, $N_1$ which would be the lightest fourth generation particle and therefore stable and $N_2$ which would decay to $N_1+Z$ \cite{Carpenter}. CDF has searched for the pair production of $N_2$, followed by the decay $N_1+Z$ \cite{nusearch}. They select events in which one $Z$ boson decays to electrons or muons and the other to jets by requiring two high-$p_T$ leptons of the same flavor, opposite charge, and a pair mass consistent with the $Z$ boson mass, two jets without $b$-tags and large missing $p_T$. CDF finds no excess beyond standard model expectations and sets upper limits for the $N_2$ pair production cross section between 50 and 1000 fb for $N_2$ masses between 175 and 350~GeV and $N_1$ masses between 75 and 225~GeV. These limits are not yet close to the predicted cross sections which are below 1 fb.

\section{Exotic Resonances}

\subsection{Resonances that decay to $t\bar{t}$}

Resonances that decay to $t\bar{t}$ final states have been predicted by many models of physics beyond the standard model. They occur in GUTs with larger symmetry groups, as Kal\-uza-Klein excitations of the gluon or the $Z$ boson, as axigluons, or as manifestations of new strong dynamics.The benchmark model for $t\bar{t}$ resonances has been a leptophobic topcolor $Z'$ with a width equal to 1.2\% of its mass that decays exclusively to the $t\bar{t}$ final state \cite{TopColorZp}.

\begin{figure}[bh]
\includegraphics[width=\columnwidth]{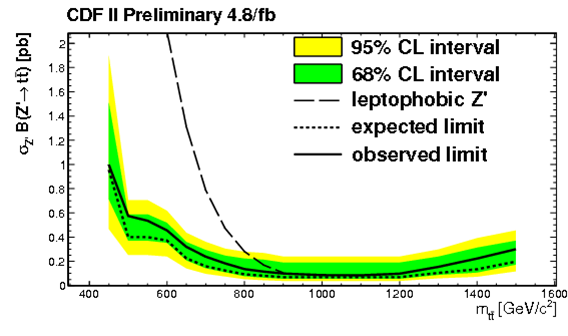}
\caption{Cross section limits for the production of a narrow $t\bar{t}$ resonance from the CDF experiment.}
\label{fig:MttlimitsCDF}       
\end{figure}

\begin{figure}
\includegraphics[width=\columnwidth]{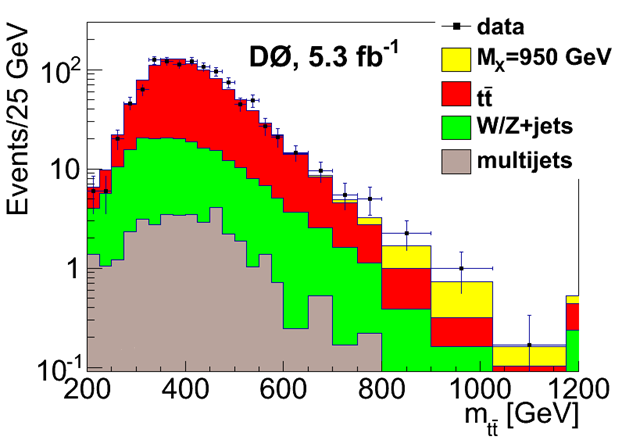}
\caption{Distribution of the reconstructed $t\bar{t}$ mass for events with at least four jets from the D0 experiment.}
\label{fig:MttD0}       

\includegraphics[width=\columnwidth]{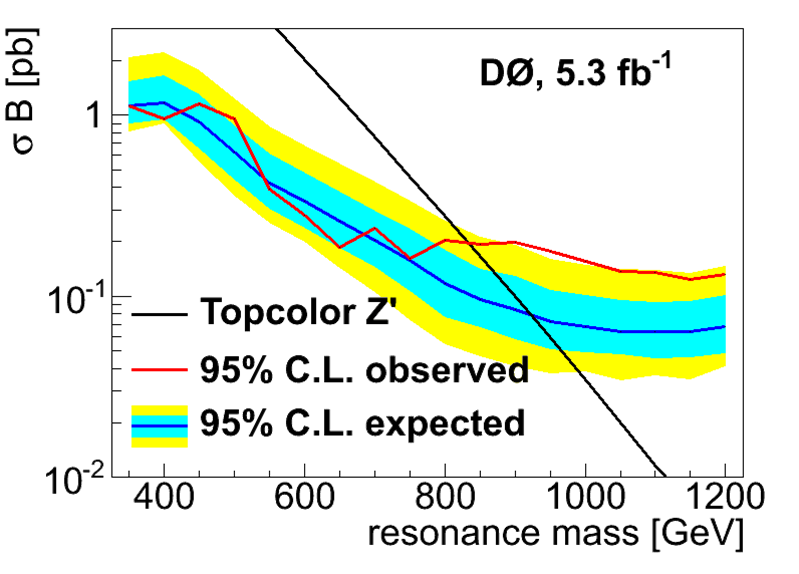}
\caption{Cross section limits for the production of a narrow $t\bar{t}$ resonance from the D0 experiment.}
\label{fig:MttlimitsD0}       
\end{figure}

CDF and D0 have searched for such resonances~\cite{ttsearches}. The event selection requires a high-$p_T$ electron or muon and missing $p_T$, a number of jets (at least four for CDF and at least three for D0), one of which is tagged as a $b$-jet. CDF computes the probability density as a function of resonance mass, obtained using the matrix elements for the processes involved. Figure~\ref{fig:MttlimitsCDF} shows the CDF limits for the production cross section of a narrow $t\bar{t}$ resonance. D0 reconstructs the invariant mass of the system consisting of the leptonically decaying $W$ boson and the leading three or four jets (Fig.~\ref{fig:MttD0}) and sets the limits shown in Fig.~\ref{fig:MttlimitsD0}. Based on the CDF and D0 results a topcolor $Z'$ boson is disfavored for masses below 900~GeV.

\subsection{Resonances that Decay to $ZZ$}

\begin{figure}[th]
\centering
\includegraphics[width=0.9\columnwidth]{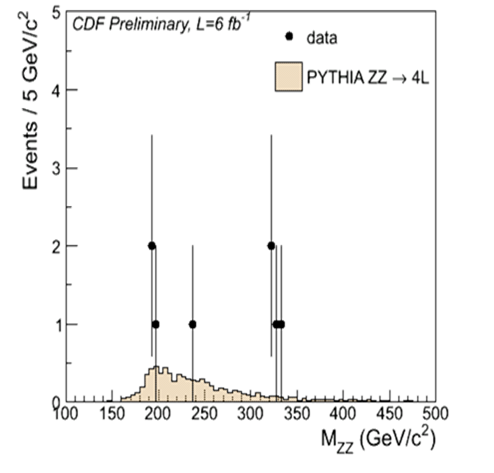}
\caption{Distribution of the $ZZ$ pair mass for four-lepton events from the CDF experiment.}
\label{fig:MZZCDF}       

\centering
\includegraphics[width=0.85\columnwidth]{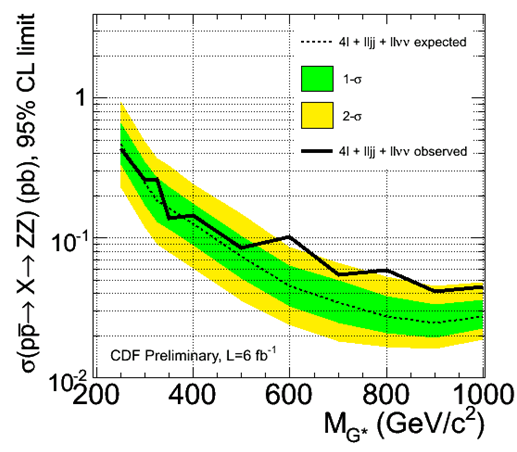}
\caption{Cross section limits for the production of $ZZ$ pairs from the CDF experiment.}
\label{fig:ZZlimits}       
\end{figure}

CDF has searched for resonances that decay to pairs of $Z$ bosons~\cite{ZZCDF}. Such resonances arise, for example, as Kaluza-Klein modes of Randall-Sundrum gravitons. The analysis is designed to search for events in which both $Z$ bosons decay to electrons or muons. The branching fraction for this decay is small - about 0.5\% but there is essentially no background for this four-lepton event signature. In an integrated luminosity of 6 fb$^{-1}$, CDF observed eight events, which is roughly twice the number of events expected from the continuum $ZZ$ production predicted by the standard model. Four of these events cluster within 7~GeV of a $ZZ$ pair mass of 327 GeV (Fig.~\ref{fig:MZZCDF}). The $p$-value for observing an excess of this size is about $1-2\times10^{-14}$. 
If this excess originates from anomalous $ZZ$ production similar excesses should appear in other $ZZ$ decay channels which have larger branching fractions but higher backgrounds. CDF has examined the channels $ZZ\rightarrow \ell^+\ell^-\nu\nu$ and $ZZ\rightarrow\ell^+\ell^-q\bar{q}$.  None of them show any excess over standard model expectations. D0 also has performed a measurement of the $ZZ$ cross section in the four-lepton channel~\cite{ZZxsecD0} and does not see any excess in its data. Based on the four-lepton data, CDF sets the limits shown in Fig.~\ref{fig:ZZlimits} for the production cross section of a narrow $ZZ$ resonance.

\section{Universal Extra Dimensions}

Additional spacetime dimensions could resolve the hierarchy problem and keep the Higgs mass small in a natural way. If there are more than four dimensions the observed Planck scale would only be an effective four-dimensional scale and the true scale of gravity could be much lower, near the electroweak scale. Extra spacetime dimensions in which all fields can propagate are called universal extra dimensions. One would then expect the existence of Kaluza-Klein  excitations for all particles with a mass spectrum $M_n^2 = M_0^2 + n^2/R^2$, where $M_n$ is the mass of the $n^{th}$ Kaluza-Klein excitation, $n=0$ corresponds to the standard model particle that has no momentum in the extra dimension, and $1/R$ is the compactification scale of the extra dimension. Current experimental limits still allow for this scale to be on order of the weak scale. This scale determines the mass spectra of the Kaluza-Klein modes~\cite{AppelquistChengDobrescu}. The number of Kaluza-Klein modes is conserved so that there cannot be any vertices with an odd number of Kaluza-Klein modes. The lightest mode that propagates in the extra dimension corresponds to $n=1$ and could be accessible at the Tevatron. The Kaluza-Klein excitations of the gluon, $g_1$, and the quarks, $q_1$, would be produced in pairs through the strong interaction at the Tevatron and they would decay to the lightest Kaluza-Klein particle, the Kaluza-Klein excitation of the photon, $\gamma_1$, through a cascade decay. In this decay leptons of the same charge are produced with a branching fraction of about 1\%. 

\begin{figure}
\includegraphics[width=\columnwidth]{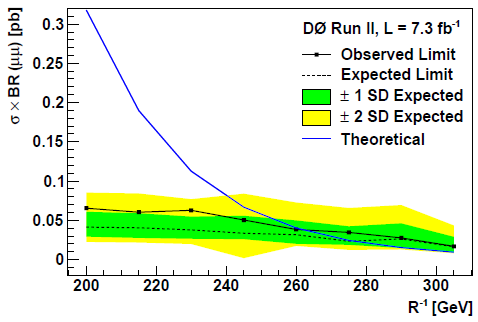}
\caption{Cross section limits for the production of like-sign dimuon events from the D0 experiment.}
\label{fig:UEDlimits}       
\end{figure}

D0 has searched for evidence for a universal extra dimension by analyzing an inclusive sample of events with same sign muons~\cite{UEDsearch}. Multijet production is a significant background for this signature and it is estimated from data. The predicted spectra of kinematic quantities agree well with observation and a boosted decision tree~\cite{BDT} was trained using many input variables to provide a discriminant for the Kaluza-Klein signal and background. The distribution of this discriminant is used to obtain the limits shown in Fig.~\ref{fig:UEDlimits}. The results disfavor compactification scales $1/R<260$~GeV and indicate that the mass of the lightest Kaluza-Klein quark must be above 317 GeV. 

\section{Dark Matter}

The astrophysical evidence for dark matter is the most compelling evidence that there must exist matter that is not accounted for in the standard model. If dark matter is made up of a new elementary particle  this particle may be produced by high energy colliders. Such a particle must be stable, implying that it carries a conserved new charge. There then has to be a "connector" particle that carries both strong charge and this new charge in order to produce the dark matter particle in collisions of standard model particles. 

\begin{figure}
\centering
\includegraphics[width=0.9\columnwidth]{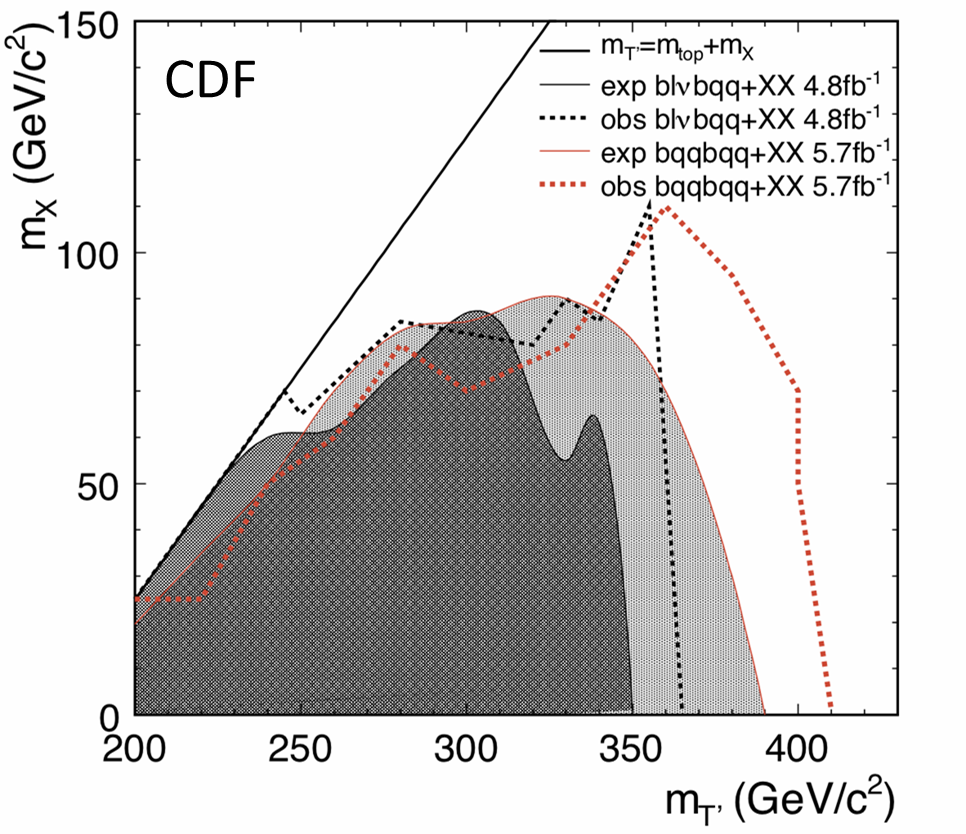}
\caption{$T'$ and $X$ masses for which the limits from CDF are below the expected cross sections. The shaded area corresponds to the observed limits, the dashed lines to the expected limits.}
\label{fig:DMlimits}       
\end{figure}

One such scenario is a top quark partner $T'$ that carries strong charge and the new charge and decays into a top quark and a dark matter particle~\cite{DM}. CDF has carried out an analysis to 
look for $T'\bar{T'}\rightarrow t\bar{t}+X\bar{X}$, where $X$ is the dark matter particle which will escape the detector unseen. Thus these events will result in the same final states as $t\bar{t}$ production except for larger missing $p_T$ from the escaping dark matter particles. CDF considered the lepton+jets channel with an electron or muon, at least four jets and missing $p_T$ in excess of 100-160~GeV, and the all hadronic channel with 5-10 jets, large $H_T$, and missing $p_T>50$~GeV~\cite{DMCDF}. No evidence for dark matter production has been observed and CDF was able to set limits on production cross sections. Figure~\ref{fig:DMlimits} shows the mass values for $T'$ and $X$ for which the cross section limit is below the expected cross section. 

\section{Conclusion}

The Fermilab Tevatron collider has delivered an integrated luminosity of 12 fb$^{-1}$ to the 
CDF and D0 experiments which continue to search for deviations from standard model predictions. Small deviations have been observed but there is no smoking gun for physics beyond the standard model.

\section*{Acknowledgements}

I would like to thank the organizers for inviting me to pres\-ent these results and for organizing a stimulating conference and my colleagues from the CDF and D0 experiments for providing me with their latest results and many useful suggestions for this presentation.

\end{document}